# A photonic crystal based broadband graphene saturable absorber

Saifeng Zhang,[1,2] Caoran Shen,[1,2] Ivan M. Kislyakov,[1,2*] Ningning Dong,[1,2]

Anton A. Ryzhov,[3] Xiaoyan Zhang,[1,2] Inna M. Belousova,[3] Jean-Michel Nunzi,[1,2,4]

and Jun Wang[1,2,5*]

[1]*Laboratory of Micro-Nano Photonic and Optoelectronic Materials and Devices, Shanghai Institute of Optics and Fine Mechanics, Chinese Academy of Sciences, Shanghai 201800, China*

[2]*Key Laboratory of Materials for High-Power Laser, Shanghai Institute of Optics and Fine Mechanics, Chinese Academy of Sciences, Shanghai 201800, China*

[3]*Institute for Laser Physics, S. I. Vavilov State Optical Institute, Saint Petersburg 199053, Russia*

[4]*Department of Physics, Engineering Physics and Astronomy and Department of Chemistry, Queen's University, Kingston K7L-3N6, Canada*

[5]*State Key Laboratory of High Field Laser Physics, Shanghai Institute of Optics and Fine Mechanics, Chinese Academy of Sciences, Shanghai 201800, China*

**Abstract**

The enhanced nonlinear optical response of a one-dimensional (1D) photonic crystal (PC) made from polymers and graphene composites is observed. The graphene PC was fabricated by spin-coating. It shows obvious bandgaps at two wavelengths in transmittance. Femtosecond Z-scan measurement at 515 nm and 1030 nm reveals a distinct enhancement in the effective nonlinear absorption coefficient $\beta_{eff}$ (and imaginary part of third-order dielectric susceptibility $Im\chi^{(3)}$) for graphene nanoflakes embedded in the PC, when compared with the bulk graphene-polymer dispersion. The effect is studied in a wide range of laser intensities. The inclusion of graphene into a 1D-PC remarkably decreases the saturable absorption threshold and saturation intensity, providing a desired solution for an advanced all-optical laser mode-locking device.

*Authors to whom correspondence should be addressed. Electronic address: jwang@siom.ac.cn, iv.kis@mail.ru

The unique mechanical, thermal, electronic and photonic features of graphene attract much attention. Single two-dimensional graphene layers have a singular structure with K-point linear dispersion relation, zero-bandgap and massless Dirac fermion carriers.[1,2] The difference from bulk graphite crystal induces unusual nonlinear optical (NLO) properties.[3,4] It can generate electron-hole pairs upon laser excitation at any wavelength λ which makes it a progressive solution to ultrafast saturable absorption (SA)[5-9] and optical limiting.[10,11] However, small NLO-coefficient values in natural materials restrict NLO applications to high intensity thresholds.

It is understood that embedding NLO materials in a photonic crystal (PC) can yield lower NLO thresholds.[12-16] The 1D-PC is one of the simplest structures. It consists of alternating dielectric materials with different optical refractive indices *n*. Due to the destructive interferences in such periodic structure, a spectral bandgap appears in the linear transmission spectrum, meaning that light with certain λ gets reflected by the PC. Bandgap width and depth are determined by the materials *n*-difference and by the total number of alternating layers.[12] If the PC embeds high NLO response materials, even weak intensity laser beams change its *n*-value due to field enhancement inside the structure,[13,16] which shifts the bandgap. Therefore, NLO threshold in a PC can be much smaller than in bulk materials, making it crystal an exciting NLO-device design.

Earlier computer simulation of an embedded graphene-monolayer 1D-PC structure showed SA-threshold decrease by two orders of magnitude for an 818 nm-pump wavelength.[17] We propose a simpler and more accessible technology for PC-manufacturing, based on the polymer dispersion of graphene flakes. We recently experimentally demonstrated the possibility of graphene nanoflakes coated on a mirror to provide intense and fast NLO response with decreased losses for 2 μm-lasers.[18] Developing on this technology, we present a graphene-containing PC designed for IR and visible wavelength pairs, suitable for the two harmonics of a standard laser source.

To prepare the crystal, graphene was first suspended using the liquid phase exfoliation method.[19] That is, graphite powder was bath-sonicated in a sodium cholate aqueous solution (0.5 wt.%) for 24 h; with 5 mg/ml graphite concentration. The suspension was then settled for 24 h and its supernatant was taken with a pipette. It was centrifugated using a ROTOFIX-32A centrifuge for 90 min at 1500 rpm and the supernatant was again taken.

The 1D-PCs under study consist of twelve alternating layers of poly(9-vinylcarbazole) (PVK) and of graphene-based poly(vinyl-alcohol) (PVA) nanocomposite, see Fig. 1(a). The periodic structure was spin coated on a glass substrate. The PVK toluene solution was spin coated for 60 s at 500 rpm, followed by 20 s at 4000 rpm. Next, the graphene suspension mixed in the PVA aqueous solution (PVA-G) was spin coated for 60 s at 500 rpm, followed by 20 s at 6000 rpm. PC-preparation by spin coating was possible owing to a combination of solvents with contrasted polarity: nonpolar solvent (toluene) for PVK versus polar solvent (water) for PVA-G. This guaranteed the integrity of each layer during the process. A PC-G1 sample picture is shown in the Fig. 1(b). The green spot in its center reveals its maximal reflectance in the green.

In order to compare the NLO response of graphene in PC, a second PC sample (PC-G2) and a pure PVA-G film were fabricated by spin-coating from different polymer concentrations. PVK concentrations varied from 10 to 30 g/l, and those of PVA – from 18 to 27 g/l. Thicknesses of PC-G1, PC-G2 and PVA-G are 1732 nm, 2961 nm and 1436 nm respectively, as measured by a Bruker-Dektak stylus-profiler. Viscosity of the polymer solutions determines the thickness differences.

Linear transmission of two samples measured with a PerkinElmer Lambda 750 dual-beam UV-VIS spectrophotometer is shown as solid lines in Fig. 1(b). It is larger than 80% throughout the visible and near-IR in pure PVA-G film.

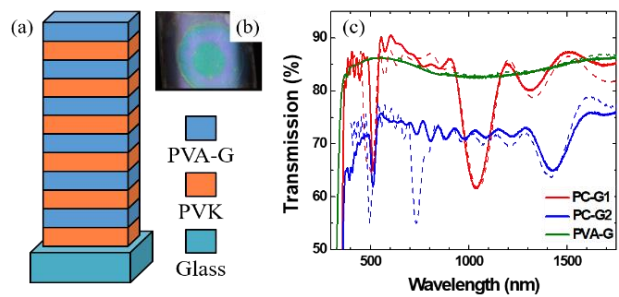

Fig. 1. (a) Structure of the spin-coated PC; (b) PC-G1 sample picture; (c) measured (solid line) and calculated (dashed line) linear transmission spectra of PC-G1, PC-G2 and PVA-G.

Two bandgaps near 515 nm and 1030 nm are seen in the PC-G1 structure. The transmission minimum (bandgaps bottom) of PC-G1 is close to 65%. The resonator Q-factors: $\nu_0/\Delta\nu$ are 7 at 1030 nm and 15 at 515 nm. The transmission minimum of PC-G2 is about 70%, and the transmission maximum is lower than those of PC-G1 and PVA-G owing to its larger thickness than the two others. Two bandgaps are seen near 515 nm and 1400 nm. At 515 nm Q-factor is 20.

Spectra of the layered structures were calculated using the matrix method[20] based on the known $n(\lambda)$ for PVA[21] and PVK.[22] The ratio between PVK and PVA-G layer thicknesses, $L_{PVK}/L_{PVA-G}$, within a pair of neighboring layers was varied in a wide range to fit the experimental spectra consistently with the known total thickness of the film. The best coincidences are shown in Fig. 1(b) and correspond to $L_{PVK}/L_{PVA-G}$ = 11/4 in PC-G1 and 3/22 in PC-G2. The second-order bandgap at 730 nm in PC-G2 is probably suppressed by layers nonuniformity (defects), $\delta L$, increasing the dephasing $\Delta\varphi = 4\pi n(L \pm \delta L)/\lambda$ of reflected beams for shorter wavelengths. However, the numerical curves reproduce the broadband parts and the main bandgap shapes in the measured spectra. Therefore, the simulations provide us with the information about layer thicknesses inside the PCs, which we assume are realistic.

NLO properties of the three samples were measured by open aperture Z-scan set up. Laser source was a fiber laser with 340-fs pulse width and 100 Hz-repetition rate. We use a 15 cm focal length lens to focus the laser beam on the sample. The transmitted signal and a reference signal were collected by fast silicon detectors. All experiments were performed at the main wavelength 1030 nm and its second harmonic 515 nm. They both perfectly match the two bandgaps of PC-G1, but only the second harmonic corresponds to a bandgap in PC-G2.

Figures 2(a) and 2(b) show the open aperture Z-scan transmissions for PC-G1, PC-G2 and PVA-G, obtained at similar input intensities in focus: 3.5, 3.7, 5.1 MW/cm$^2$ at 515 nm and 2.9, 3.3, 3.4 MW/cm$^2$ at 1030 nm respectively. All curves show more pronounced SA behavior for the PCs than for the non-crystalline film at both wavelengths.

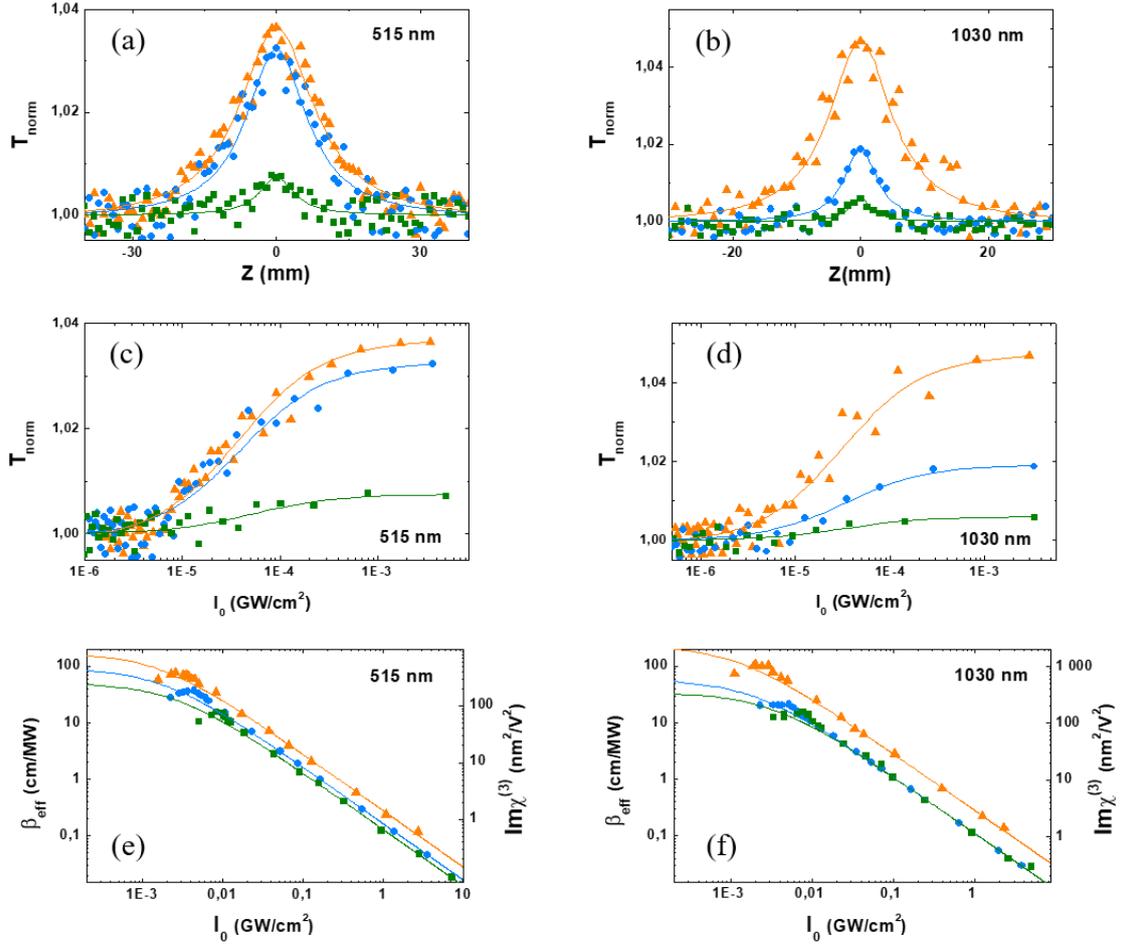

Fig. 2. (a) and (b) open-aperture Z-scan data at $I_0 \approx 4$ MW/cm$^2$; (c) and (d) normalized transmission *vs.* input intensity; (e) and (f) $\beta_{eff}$ and Im$\chi^{(3)}$ respectively *vs.* input intensity. Points: (▲) PC-G1; (●) PC-G2; (■) PVA-G; curves: (a) – (d) fitting by Eq. (1), (e) and (f) fitting by Eq. (3).

The observed peculiarity of the PC-G1 structure is its larger bleaching contrast at 1030 nm than at 515 nm (4.7% against 3.6%), that follows the bandgap depth: more profound in the IR range. The SA-contrast of PC-G2 is inverse: 1.9% at 1030 nm and 3.2% at 515 nm, due to the absence of a bandgap near 1030 nm. The similar contrast of 0.6%-0.7% in PVA-G also correlates with the bandgap absence.

The NLO response was quantified by the effective nonlinear coefficient $\beta_{eff}$ obtained from fitting the Z-scan curves in Figs. 2(a) and 2(b) according to the formula:

$$T_{Norm}(z) = \ln\left[\left(1+\beta_{eff}I(z)L_{eff}\right)/\left(\beta_{eff}I(z)L_{eff}\right)\right], \quad L_{eff} = \left[1-\exp(-\alpha_0 L)\right]/\alpha_0, \quad (1)$$

where $I(z)$ is the input intensity with the value $I_0$ at focus:

$$I(z) = I_0 / \left[1+(z/z_0)^2\right], \quad (2)$$

$\alpha_0$ is the linear absorption coefficient, and $z_0$ is the Rayleigh length.

To determine SA-thresholds, we plotted the normalized transmission against incident laser intensity, calculated from z position using Eq. (2), in which $z_0$ is evaluated through the beam waist radii: $z_0 = \pi w_0^2/\lambda$, $w_0 = 23$ μm (at 515 nm) and 45 μm (at 1030 nm). The resulting curves, corresponding to those from Figs. 2(a) and 2(b) are shown in Figs. 2(c) and 2(d). In this scale the initial changes of transmittance are more observable, and the SA-threshold can be read as the intensity where the transmission (the fitting curve as the average property) overgrows 10% of its contrast.

We Z-scanned our samples at different focus intensities: from $10^{-3}$ to 10 GW/cm$^2$, every time evaluating $\beta_{eff}$ and the corresponding imaginary part of the third-order nonlinear susceptibility: $\mathrm{Im}\chi^{(3)} = \left(\varepsilon_0 c \lambda n^2 \beta_{eff}\right)/(6\pi)$, where $\varepsilon_0$ is vacuum permittivity. Results are presented in Figs. 2(e) and 2(f). At high intensities the data points follow the homogeneously broadened saturable absorption model:

$$\alpha_{sat} = \alpha_0/(1+I/I_{sat}), \quad \beta_{eff} = \alpha_0/(I_{sat}+I), \qquad (3)$$

represented in the figures by solid curves. The fitting with this model provided the saturation intensity, $I_{sat}$, and the graphene effective linear absorption coefficient, $\alpha_{0,eff}$, which is hardly evaluated based on the transmission spectra because of the mirror structure of the samples. We can see that at low intensities $\beta_{eff}$ is smaller and the excited states are saturated faster than it is predicted by the model. It may indicate an influence of other processes which reduce the bleaching, *e.g.* two-photon absorption like it was demonstrated for an epitaxial bilayer graphene.[23]

The SA parameters obtained based on Z-scan fitting by Eq. (1) and $\beta_{eff}$ (I) by Eq. (3) are presented in Table 1. Their values for the bulk film fall within the significant range of variation observed for PVA-G films prepared by different methods.[6,23-26] This indicates a strong dependence of SA on thickness and quality of the graphene sheets. We can also see that in all aspects SA is several times larger in PCs than in the bulk film at both wavelengths: $\beta_{eff}^{PC}/\beta_{eff}^{film} \approx 2\text{-}9$, $I_{thr}^{film}/I_{thr}^{PC} \approx 1.5\text{-}3$, $I_{sat}^{film}/I_{sat}^{PC} \approx 1.4\text{-}2.7$. This demonstrates that the PC structure successfully enhances the NLO response of graphene at similar input intensity, which can be explained by photon localization (larger Q-factor) in the PC.[12]

Table 1. Saturation parameters obtained from Z-scan at moderate intensity: $\beta_{eff}$, $z_0$, $I_{thr}$, F, and from intensity dependence of $\beta_{eff}$: $\alpha_{0,eff}$, $I_{sat}$.

| Structure | $\beta_{eff}$, cm/MW | $I_{thr}$, kW/cm$^2$ | $\alpha_{0,eff}$, cm$^{-1}$ | $I_{sat}$, MW/cm$^2$ | F | $\beta_{eff}$, cm/MW | $I_{thr}$, kW/cm$^2$ | $\alpha_{0,eff}$, cm$^{-1}$ | $I_{sat}$, MW/cm$^2$ | F |
|---|---|---|---|---|---|---|---|---|---|---|
| | 515 nm | | | | | 1030 nm | | | | |
| PC-G1 | -71.1 | 4 | 280 | 1.6 | 6.8 | -102.0 | 2 | 280 | 1.2 | 8.2 |
| PC-G2 | -35.2 | 4 | 165 | 1.8 | 3.4 | -20.6 | 5 | 110 | 1.9 | 1.7 |
| PVA-G | -10.5 | 10 | 130 | 2.6 | - | -12.4 | 6 | 110 | 3.2 | - |

The enhanced NLO response is clearly evidenced through an enhancement factor, F, determined as the ratio of PC-to-bulk $\beta_{eff}$ values. For a wide bandgap like that at 1030 nm, the enhancement is apparently due to photon localization since the difference in F between resonating PC-1 and nonresonating PC-2 is significant. But in the visible range the optical Kerr-nonlinearity $n_2 I_0$ can give a remarkable shift of the resonance wavelength: $\Delta\lambda = 2L_{PVA-G} n_2 I_0$. We can roughly upper estimate this shift based on the known $Re\chi^{(3)}_{1l} = 1.5\cdot10^{-7}$ esu of a single graphene layer[27] and the effective number of graphene layers $N_{eff} = \alpha_{0,eff} L_{PVA-G}/(\sigma_{Gr}\rho_s)$, where $\sigma_{Gr}$ = 8000 cm$^2$/g is the absorption cross-section of our graphene suspension[28] and $\rho_s$ = 77 ng/cm$^2$ is the graphene surface density. Using $n_2 = 12\pi^2 10^7 Re\chi^{(3)}_{1l} N_{eff}/(n^2 c)$, we obtain $n_2$ = 0.010 cm$^2$/MW in PC-G1 and 0.033 cm$^2$/MW in PC-G2. At intensity $I_0$ = 3.5 MW/cm$^2$ it can give $\Delta\lambda$ = 6 nm for PC-G1 and 100 nm for PC-G2. Despite a high Q-factor this shift reduces the field enhancement effect in PC-G2 and determines its bleaching through the bandgap depth which is half that in PC-G1.

In conclusion, we fabricated the 1D-PC with alternating layers of PVA-G and PVK films to enhance the NLO response of graphene. Graphene nanosheets were exfoliated and dispersed in the polymer solution, and the PC was constructed by spin-coating. Periodic structures with two bandgaps at visible and IR wavelengths were obtained. Significant enhancement of the nonlinear absorption coefficient $\beta_{eff}$ and the imaginary part of the third-order nonlinear susceptibility $Im\chi^{(3)}$ was obtained in the PC structures as compared to the bulk PVA-G film. We also showed a remarkable decrease of SA-threshold and saturation intensities of graphene embedded in the PC in comparison with bulk PVA-G. SA enhancement factor reaches 7 in the visible and 8 in the IR. Therefore, the 1D-PC design can enhance SA of materials embedded in a periodic structure. The combination of graphene, polymer compounds, layer number and thicknesses can be optimized to yield highly effective NLO devices.

**Acknowledgements**

This work is supported by National Natural Science Foundation of China (NSFC) (61675217, 61522510), Strategic Priority Research Program of Chinese Academy of Sciences (CAS) (XDB16030700), Key Research Program of Frontier Science of CAS (QYZDB-SSW-JSC041), Program of Shanghai Academic Research Leader (17XD1403900), Natural Science Foundation of Shanghai (No. 18ZR1444700), Youth Innovation Promotion Association and President's International Fellowship Initiative of CAS (2017VTA0010, 2017VTB0006, 2018VTB0007).